\documentclass[english,aps,prl,twocolumn, superscriptaddress]{revtex4}%
\usepackage[T1]{fontenc}
\usepackage[latin9]{inputenc}
\usepackage{amsmath}
\usepackage{amssymb}
\usepackage{babel}
\usepackage{amsfonts}
\usepackage{graphicx}
\usepackage{float}
\usepackage{physics}
\usepackage{xcolor}

\begin{document}
\title{Comment on ``Observing the "quantum Cheshire cat" effect with noninvasive weak
measurement'' }
\author{Surya Narayan Sahoo}
\affiliation{Light and Matter Physics, Raman Research Institute, Bengaluru 560080, India
}
\author{Dipankar Home}
\affiliation{Center for Astroparticle Physics and Space Science (CAPSS), Bose Institute,
Kolkata 700 091, India}
\author{Alex Matzkin}
\affiliation{Laboratoire de Physique Th\'eorique et Mod\'elisation, CNRS Unit\'e 8089, CY
Cergy Paris Universit\'e, 95302 Cergy-Pontoise cedex, France}
\author{Urbasi Sinha}
\email{usinha@rri.res.in}
\affiliation{Light and Matter Physics, Raman Research Institute, Bengaluru 560080, India
}

\begin{abstract}
In a very recent work [arXiv:2004.07451], Kim et al claimed to have made the
first genuine experimental observation of the Quantum Cheshire Cat effect. We
dispute this claim on the ground that the setup employed is not adequate for
making the weak measurements that define this interferometric effect. Half of
the necessary weak values are not observed, and the other half is obtained
indirectly by combining results measured with distinct setups.

\end{abstract}
\maketitle

In their work \cite{kim}, henceforth designated by ``P'', Kim et al. report what
they term "\emph{the first genuine experimental observation of the quantum
Cheshire cat effect}". In this Comment, we argue that their claim does not
withstand scrutiny. Our main argument is that the setup employed in P is
inadequate to observe the quantum Cheshire cat (QCC) effect because the very
nature of the effect requires weak measurements (WM) to be carried out on both
arms of the interferometer.\ In P however WM are made on a single arm;
moreover these WM are not the ones relevant for the QCC effect.\ Rather the WM
required for observing the QCC are deduced from independent WM of a different
observable.

\begin{figure}[htb]
	\includegraphics[width=0.99\linewidth]{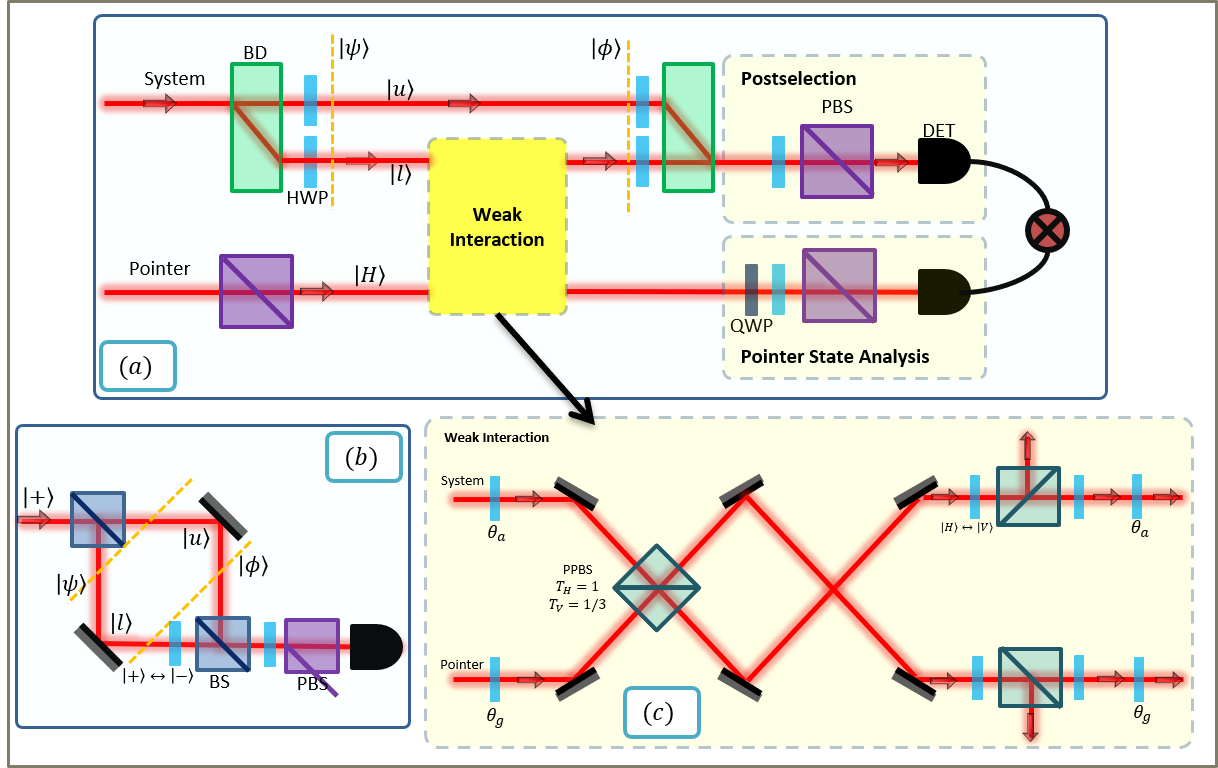}\caption{$[a]$The experimental
		setup used in P. The two beam displacers (BD) form the Mach-Zehnder
		interferometer. The HWPs placed in the upper arm $\ket{u}$ and lower arm
		$\ket{l}$ makes the pre-selected state $\ket{\psi} = \frac{1}{\sqrt{2}}
		(\ket{u}\ket{+} + \ket{l} \ket{-})$ and the post-selected state $\frac
		{1}{\sqrt{2}}(\ket{u}+\ket{l})\ket{+}$. The pointer state is the polarization
		degree of freedom of another photon which has to be prepared in $\ket{H}$.
		$[c]$ The weak interaction is implemented by small rotation of $C_{z}$ gate
		effectively achieved by two-photon interference at partially polarizing beam
		splitters (PPBS) and subsequent amplitude adjustments of the output system and
		pointer states. The implementation of weak interaction is subject to
		coincidence of system and pointer photons later which occurs with a
		probability of $1/9$. $[b]$ The Mach-Zehnder interferometer for QCC
		\cite{aharonov} with pre-selected state $\ket{\psi} = \frac{1}{2}%
		(\ket{u}+\ket{l})\ket{+}$ and post-selected state $\ket{\phi} = \frac{1}%
		{2}(\ket{u}+\ket{l})\ket{+}$.}%
\end{figure}

The QCC effect \cite{aharonov} describes the spatial separation of a particle
and one of its properties in the sense of WM. A single photon traveling in a
Mach-Zehnder type interferometer has a weak value of $1$ for the spatial
projector $\Pi_{u}$ on the upper arm \emph{u} and $0$ on the lower arm
\emph{l}; for a given property, say linear polarization represented by
$\sigma_{z}$, the opposite holds.\ Hence the weak values (WV) defining the QCC
are%
\begin{align}
\left\langle \Pi_{u}\right\rangle _{w}  &  =1\qquad\left\langle \Pi_{u}%
\sigma_{z}\right\rangle _{w}=0\label{d1}\\
\left\langle \Pi_{l}\right\rangle _{w}  &  =0\qquad\left\langle \Pi_{l}%
\sigma_{z}\right\rangle _{w}=1. \label{d2}%
\end{align}
All these WV can in principle be observed jointly by introducing weak
couplings between the relevant probes on each arm and the photon prepared and
detected in suitably chosen pre and post-selected states.

As remarked in P, early experiments on the QCC \cite{neutron,Qoptiq} were
unable to implement the required weak measurements and resorted instead to
strong couplings.\ Each WV in Eqs. (\ref{d1})-(\ref{d2}) was obtained
indirectly by combining several strong measurements, employing distinct
setups. This prevents the QCC from being observed because a strong measurement
aimed at extracting a given WV disturbs the other ones, so that Eqs.
(\ref{d1})-(\ref{d2}) cannot be measured jointly even in principle
\cite{annals}. Unfortunately, the setup used in P is also unable in principle
to observe the QCC.\ There are several serious experimental shortcomings as well.

The main problem is that joint measurements of the observables are not performed. Since with parametric down conversion only one photon can be used as a pointer, the system and pointer can interact only in one arm in a given run. In P,  measurements on only arm $l$ are performed.\ No
observation is done on arm $u$, so that the weak values in Eq. (\ref{d1})
cannot be determined directly from the experiment. How can one then observe spatial
separation? The authors argue (above Eq. (2) of P) that they can infer the WV on arm $u$ from complementarity.\ But to the best of our knowledge, complementarity is not well defined for WM. For strong measurements, complementarity would be applicable: then the operators in each of the sets
$\left\{  \Pi_{u},\Pi_{l}\right\}  $ and $\left\{  \Pi_{u}\sigma_{z},\Pi
_{l}\sigma_{z}\right\}  $ are complementary, but both sets cannot be measured
jointly with strong interactions. Crucially, for the measurements of $\left\{
\Pi_{u},\Pi_{l}\sigma_{z}\right\}  $ that lie at the heart of the QCC effect
(particle observed on arm $u$ and polarization on $l$) there is no such
complementarity: a joint measurement needs to be performed and this is only
possible with WM. Hence the sole measurement of WV along $l$ is in principle
insufficient in order to observe the spatial separation of properties. Note that the \textquotedblleft sum
rule\textquotedblright\ for WV could in principle have been invoked but would
have required an additional measurement: for example for the polarization the
sum rule yields \cite{jpa} $\left\langle \Pi_{u}%
\sigma_{z}\right\rangle _{w}=\left\langle \sigma_{z}\right\rangle
_{w}-\left\langle \Pi_{l}\sigma_{z}\right\rangle _{w}$ where $\left\langle
\sigma_{z}\right\rangle _{w}$ is the polarization weak value that should be
measured before the entrance or after the exit beam displacers.

A second major problem with the setup employed in P is that along arm $l$,
none of the WV of Eq. (\ref{d2}) are directly measured. Instead $\left\langle
\Pi_{l}\right\rangle _{w}$ and $\left\langle \Pi_{l}\sigma_{z}\right\rangle
_{w}$ are reconstructed from independent WM of the polarization projectors
$\Pi_{H}$ and $\Pi_{V}$. The WV (\ref{d2}) are extracted by combining the
pointer averages $\left\langle \sigma_{x}^{H}\right\rangle _{p}\pm\left\langle
\sigma_{x}^{V}\right\rangle _{p}$ obtained employing different setups (the HWP
angles $\theta_{a}$ and $\theta_{g}$ on Fig. 1(c) are changed) on different
photons. Hence even if one wanted to rely on the sum rule (by performing the
required additional measurement), it would be hard to claim that single photon
spatial separation can be observed with the present setup.

Moreover genuine weak measurements are not performed.\ In the quantum
non-demolition scheme employed in P the meter photon is not a von Neumann
pointer but is coupled to the measured photon through a nondeterministic
polarization-dependent coupling scheme \cite{pryde}. In P, the required
measurements are not done in the weak regime.\ Instead, the WV are obtained by
interpolating, from measurements made at varying strengths $g$, a quantity that should in principle be measured in the neighbourhood of $g\approx0$.\ This creates large
uncertainties and estimated WV dependent to a large extent on the chosen polynomial function and data range included in the fit (this is particularly
striking in Fig.\ 3(a) of P). 

Note further that as a result of the nondeterministic coupling scheme
employed, in which coincidences of system and pointer photons are needed, one
is actually dealing with a sub-ensemble of the original pre and post selection
ensemble. Due to the probabilistic nature of system and pointer photons
emerging in different ports after the two-photon interference and subsequent
amplitude reduction (see Fig. \ 1(c)), the overall probability of successful
implementation of weak interaction is $1/9$. Thus the experiment loses $8/9$
of the system photons that are part of the pre and post-selection ensemble.

Overall we have argued that strictly speaking the experiment reported in P
deals with interpolating WV from orthogonal polarization projector
measurements made in the sole lower arm of the interferometer. These results
represent a progress relative to the early experiments on the QCC \cite{neutron,Qoptiq}
in that weak values are obtained from quantum non-demolition measurements. But
contrary to what is claimed in P, this is still far from representing a
genuine and unambiguous observation of the Quantum Cheshire Cat effect.

\end{document}